\documentstyle[12pt]{article}

\begin{document}
\thispagestyle{empty}

\begin{center}
\LARGE \tt \bf{Hawking radiation in moving plasmas}
\end{center}
\vspace{1cm}
\begin{center} {\large L.C. Garcia de Andrade\footnote{Departamento de
F\'{\i}sica Te\'{o}rica - Instituto de F\'{\i}sica - UERJ
Rua S\~{a}o Fco. Xavier 524, Rio de Janeiro, RJ
Maracan\~{a}, CEP:20550-003 , Brasil.
E-Mail.: garcia@dft.if.uerj.br}}
\end{center}
\vspace{1.0cm}
\begin{abstract}
Bi-metricity and Hawking radiation are exhibit in non-relativistic moving magnetohydrodynamics (MHD) plasma medium generating two Riemannian effective spacetimes. The first metric is a flat metric although the speed of "light" is given by a time dependent signal where no Hawking radiation or effective black holes are displayed. This metric comes from a wave equation which the scalar function comes from the scalar potential of the background velocity of the fluid and depends on the perturbation of the magnetic background field. The second metric is an effective spacetime metric which comes from the perturbation of the background MHD fluid. This Riemann metric exhibits a horizon and Hawking radiation which can be expressed in terms of the background constant magnetic field. The effective velocity is given Alfven wave velocity of plasma physics. The effective black hole found here is analogous to the optical black hole in moving dielectrics found by De Lorenci et al [Phys. Rev. D (2003)] where bi-metricity and Hawking radiation in terms of the electric field are found.

\end{abstract}
\vspace{1.0cm}
       
\begin{center}
\end{center}

\newpage
\pagestyle{myheadings}
\markright{\underline{Effective geometry in MHD}}

\section{Introduction}
Earlier De Lorenci et al \cite{1} have shown that optical black holes descrided by Leonhardt and Pwinicki \cite{2} also exists in dielectric moving media. In their model two distinct optical metrics are found with consequent birefrigence and two respective horizons and optical black holes. Also the Hawking temperature \cite{2} is also found in terms of the electric field. Recently other effective black holes in  dielectrics have also been found by Schuetzhold et al \cite{3}. In this report we show that effective black holes \cite{2} can be found in moving plasmas from the wave equation in magnetohydrodynamics (MHD) flows. In our case although bi-metricity is found one of the effective metrics is flat with time dependent sonic speeds however this only vanishes in the case the perturbation of the magnetic field vanishes. Actually this metric is obtained from the wave equation obtained from the wave scalar which is the potential of the background velocity ${\psi}_{0}$. Nevertheless from the other wave equation of the moving plasma given for the scalar potential of the background potential of the perturbed flow ${\psi}_{1}$. Actually we have two flat Minkowski metrics. One is the laboratory metric while the other is the flat effective metric of the background fluid, besides the perturbed metric. Therefore two analogue special relativistic metrics appears in the system. The metrics are not called acoustic here neither the black holes are called sonic because in fact Alfven waves decoupled from the acoustic waves in the plasma medium. The magnetic background field is chosen to be constant \cite{4}. This paper is organised as follows: In the section 2 we address the MHD equations and its respective wave equarion with the corresponding effective MHD flat metric and its Riemannian bimetric effective geometry. In section 3 we show that Hawking radiation \cite{5} of the MHD flow  is computed from the effective black hole. In section 4 the conclusions and discussions are presented.
\section{Riemannian geometry of MHD analogue black  hole}     
Let us consider the MHD field equations of a non-dissipative perfect magnetohydrodynamical gas as
\begin{equation}
\vec{E}+\frac{1}{c}\vec{v}{\times}\vec{B}=\vec{0}
\label{1}
\end{equation}
\begin{equation} 
\frac{{\partial}\vec{B}}{{\partial}t}= {\nabla}{\times}(\vec{v}{\times}\vec{B})
\label{2}
\end{equation}
\begin{equation}
{\nabla}.\vec{B}=0
\label{3}
\end{equation}
\begin{equation}
\vec{J}=c{\nabla}{\times}\vec{H}
\label{4}
\end{equation}
\begin{equation}
{\nabla}.\vec{J}=0
\label{5}
\end{equation}
\begin{equation}
\frac{{\partial}{\rho}}{{\partial}t}+{\nabla}.({\rho}\vec{v})=0
\label{6}
\end{equation}
\begin{equation}
{\rho}\dot{\vec{v}}= -{\nabla}({\pi}+\frac{1}{2}{\mu}_{1}H^{2})-{\rho}{\nabla}g+{\mu}(\vec{H}.{\nabla})\vec{H}
\label{7}
\end{equation} 
\begin{equation}
\dot{\eta}=0 
\label{8}
\end{equation} 
where $\vec{E}$, $\vec{B}$ are the electric and magnetic induction  fields respectively while , $\vec{H}$ is the  magnetic field and ${\mu}^{*}:= 2{\mu}-{\mu}^{2}$ ,$\vec{J}$ is the electric current, connected by the relation $\vec{B}={\mu}\vec{H}$ where ${\mu}$ is the magnetic permeability and finally ${\rho}$ is the fluid density. In the paper ${\pi}$ represents the pressure with in the polytropic gas \cite{4} we have 
\begin{equation} 
{\pi}=A({\eta}){\rho}^{\gamma}
\label{9}
\end{equation}
Here ${\gamma}$ represents the adiabatic exponent of the polytropic gas. In the approximation considered here we can consider that $A({\eta})$ is approximately equal to one since this is given by \cite{4}
\begin{equation} 
A({\eta})=(\frac{{\pi}_{0}}{{\rho}_{0}{\gamma}})exp[\frac{({\eta}-{\eta}_{0})}{c_{v}}]
\label{10}
\end{equation}
where $c_{v}$ is the specific heat at constant volume. Here the magnetic diffusity ${\eta}^{B}$ vanishes and the magnetic Reynolds $R^{B}$ number is infinity accordingly with the expression
\begin{equation} 
R^{B}= \frac{|{\nabla}{\times}(\vec{v}{\times}\vec{B})|}{{\eta}^{B}|{\nabla}^{2}\vec{B}|}
\label{11}
\end{equation}
All MHD discussed here are nonrelativistic according to the approximation $\frac{|\vec{v}|}{c}<<1$. The next step is to linearize these MHD equations accordingly with the perturbation expressions
\begin{equation}
\vec{H}=\vec{H}_{0}+\vec{h}
\label{12}
\end{equation} 
\begin{equation}
{\pi}={\pi}_{0}+{\pi}_{1}
\label{13}
\end{equation}
\begin{equation}
{\rho}={\rho}_{0}+{\rho}_{1}
\label{14}
\end{equation}
\begin{equation}
{\vec{v}}={\vec{v}}_{0}+{\vec{v}}_{1}
\label{15}
\end{equation}
where ${\rho}_{1}<<{\rho}$, $|\vec{H}_{1}|<<|\vec{H}_{0}|$ and $|\vec{v}_{1}|<<|\vec{v}_{0}|$ carachterize the perturbations. Here the quantity , $\vec{v}=\vec{v_{0}}+ \vec{v}_{1}$, ${\rho}_{0}=constant$, ${\pi}_{0}$ and $\vec{H}_{0}=constant$ represent a steady-state uniform solution of the MHD equations. Thus perturbing this solution we obtain the following equations
\begin{equation}
\frac{{\partial}{\rho}_{1}}{{\partial}t}+{\nabla}.({\rho}_{0}\nabla{\Psi}_{1})+{\nabla}.({\rho}_{1}\nabla{\Psi}_{0})=0
\label{16}
\end{equation}
\begin{equation}
{\rho}_{0}\dot{\vec{v}_{1}
}+{\rho}_{1}\dot{\vec{v}_{0}}= -{\nabla}({\pi}_{1}+\frac{1}{2}{\mu}_{1}\vec{h}.\vec{H}_{0})
\label{17}
\end{equation} 
\begin{equation}
\frac{{\partial}\vec{h}}{{\partial}t}=(\vec{H}_{0}.{\nabla})\vec{v}_{1}-\vec{H}_{0}({\nabla}.\vec{v}_{1})+(\vec{h}.{\nabla})\vec{v}_{0}-\vec{h}({\nabla}.\vec{v}_{0})
\label{18}
\end{equation} 
Here we have chosen a irrotational perturbation where $v_{1}={\nabla}{\psi}_{1}$ and $v_{0}={\nabla}{\psi}_{0}$. From this expression and equation (\ref{17}) we obtain
\begin{equation}
{\rho}_{0}\frac{{\partial}}{{\partial}t}{\psi}_{1}+{\rho}_{1}\frac{{\partial}}{{\partial}t}{\psi}_{0}= -({\pi}_{1}+\frac{1}{2}{\mu}_{1}\vec{h}.\vec{H}_{0})
\label{19}
\end{equation}
Taking the time derivative again of equation (\ref{19}) yields
\begin{equation}
{\rho}_{1}\frac{{\partial}^{2}}{{\partial}t^{2}}{\psi}_{1}
+{\rho}_{0}\frac{{\partial}^{2}}{{\partial}t^{2}}{\psi}_{1}
= -({\gamma}{{\rho}_{1}}^{{\gamma}-1}\frac{{\partial}}{{\partial}t}{{\rho}_{1}}+\frac{1}{2}{\mu}_{1}[\frac{\partial}{{\partial}t}
\vec{h}].\vec{H}_{0})
\label{20}
\end{equation} 
where we have used the relation (\ref{9}) to simplify matters. Note that the first term on the RHS of the equation (\ref{19}) must be  dropped because is not a first order term in ${\rho}_{1}$. Substitution of equations (\ref{16}) and (\ref{18}) into (\ref{20}) yields
\begin{equation}
(\frac{{\partial}}{{\partial}t}+{\nabla}.\vec{v_{0}})\frac{{\rho}_{0}}{{c_{Pl}}^{2}}(\frac{{\partial}}{{\partial}t}+{\vec{v_{0}}}.{\nabla}){\psi}_{1}-{\nabla}.({\rho}_{0}{\nabla}{\psi}_{1})=\frac{{\rho}_{0}}{{\rho}_{1}} {\Box}{\psi}_{0}
\label{21}
\end{equation}
This is the usual scalar wave equation that appears in the analog gravity. In the equation (\ref{21}) the speed $c_{Pl}$ represents the plasma wave velocity given by
\begin{equation}
{c_{Pl}}^{2}= (1-\frac{{\mu}^{*}{H_{0}}^{2}}{{\rho}_{0}})
\label{22}
\end{equation}
The LHS of equation (\ref{21}) yields the effective black hole metric
\begin{equation}
\sqrt{-g}{g^{+}}^{00}= {{\rho}_{0}}{{c_{Pl}}^{2}}
\label{23}
\end{equation}
\begin{equation}
\sqrt{-g}{g^{+}}^{0j}= {{\rho}_{0}}{{c_{Pl}}^{2}}({{\vec{v_{0}}}^{T}})_{0j}
\label{24}
\end{equation}
Note that $(i,j=1,2,3)$ while $({\mu}=0,1,2,3)$, and the speed of propagating of the ${\Box}$ D'Alembertian is given by
\begin{equation}
c^{2}=[1-\frac{{\mu}^{*}\vec{h}.\vec{H_{0}}}{{\rho}_{1}}
\label{25}
\end{equation}
Note that the plus upper sign on the metric components represent that there is another metric component given by ${g^{-}}_{{\mu}{\nu}}$  given by
\begin{equation}
{g^{-}}^{ij}= \frac{{\rho}_{0}}{c}{\delta}_{ij}
\label{24}
\end{equation}
where $(i,j=1,2,3)$ and the plasma effective spacetime speed $c_{P}$ is given by
\begin{equation}
{{c_{Pl}}^{2}}=[\frac{{\mu}_{1}\vec{H^{2}}_{0}}{{{\rho}_{0}}^{2}}]
\label{25}
\end{equation}
The plasma effective metric is 
\begin{equation}
ds^{2}=\frac{{\rho}_{0}}{c_{Pl}}[{c_{Pl}}^{2}dt^{2}-{\delta}_{ij}(dx^{i}-{v_{0}}^{i}dt)(dx^{j}-{v_{0}}^{j}dt)]
\label{26}
\end{equation}
Therefore the MHD effective Riemannian geometry ${g^{-}}_{{\mu}{\nu}}$ can be described by the special relativity effective line element
\begin{equation}
ds^{2}=\frac{{\rho}_{0}}{c}[c^{2}dt^{2}-{\delta}_{ij}dx^{i}dx^{j}]
\label{27}
\end{equation}
However in this case the signal velocity is not constant.
\section{Hawking radiation and horizons in moving plasmas}
In this section we analyze the issue of the existence of Hawking radiation and horizons in the first metric ${g^{+}}_{{\mu}{\nu}}$. From the expression    
\begin{equation} 
{g^{+}}_{00}={c_{Pl}}^{2}-{v_{0}}^{2}
\label{28}
\end{equation}
we obtain the horizon as the usual black hole condition in general relativity $g_{00}=0$ which in our case yields
\begin{equation} 
{c_{Pl}}^{2}={v_{h}}^{2}=1-\frac{{\mu}^{*}{H_{0}}^{2}}{{\rho}_{0}}
\label{29}
\end{equation}
to better comparer with MHD equation we write down in three space as
\begin{equation} 
{\nabla}^{2}{\Psi}_{1}+ \vec{K}.{\nabla}{\Psi}_{1}=0
\label{28}
\end{equation}
By the definition of Hawking radiation \cite{6}
\begin{equation}
kT_{H}= \frac{\bar{h}}{2{\pi}}\frac{g_{H}}{c_{H}}
\label{29}
\end{equation}
where $g_{H}$ is the surface gravity defined by
\begin{equation}
g_{H}=\frac{1}{2}\frac{d}{dr}[{c_{Pl}}^{2}-{v_{0}}^{2}(r,t)]
\label{30}
\end{equation}
From these expressions we obtain the form of Hawking radiation
\begin{equation}
kT_{H}= \frac{\bar{h}}{2{\pi}}\frac{g_{H}}{1-\frac{{\mu}^{*}{H_{0}}^{2}}{{\rho}_{0}}}
\label{31}
\end{equation}
This result is similar  with that one obtained by De Lorenci et al \cite{1}. 

\section{Conclusions}
Hawking radiation once more proves to be a radiation independent of the physical system or the curved spacetime of general relativity, being computed in moving plasmas as it was done previously in other physical systems such as dielectrics and viscous fluids. Bimetricity in the plasma medium is obtained.

\section*{Acknowledgments}
\paragraph*{}

I am very much indebt to Bill Unruh, Ralf Schuetzhold, Stephano Liberati,Santiago Bergliaffa , P. S. Letelier and C. Furtado for enlightening discussions on the subject of this paper. Financial supports from CNPq. (Brazilian Government Agency) and Universidade do Estado do Rio de Janeiro (UERJ) are gratefully acknowledged.

\end{document}